 \documentclass[prc,amsmath,amsfonts,showpacs,eqsecnum,floatfix,twocolumn]{revtex4}
 \usepackage{graphicx}  

\newcommand{\beq}{\begin{eqnarray}}
\newcommand{\eeq}{\end{eqnarray}}

\begin{document}

%\title{Can be recognized impact of hyperon mixing in neutron stars by accurate measurements
%of their radii ?}
\title{Quark phases in neutron stars consistent with implications of NICER}

\author{Y.\ Yamamoto$^{1}$}
\email{yamamoto@tsuru.ac.jp}
\author{N.\ Yasutake$^{2}$}
\author{Th.A.\ Rijken$^{3}$$^{1}$}
%\author{D.\ Blaschke$^{4}$$^{5}$}
\affiliation{
$^{1}$RIKEN Nishina Center, 2-1 Hirosawa, Wako, 
Saitama 351-0198, Japan\\
%Nishina Center for Accelerator-Based Science,
%Institute for Physical and Chemical
%Research (RIKEN), Wako, Saitama, 351-0198, Japan\\
$^{2}$Department of Physics, Chiba Institute of Technology, 2-1-1 Shibazono
Narashino, Chiba 275-0023, Japan\\
$^{3}$IMAPP, Radboud University, Nijmegen, The Netherlands
}

%\date{\today}
%

\begin{abstract}
The analyses for the NICER data imply 
$R_{2.0M_\odot}=12.41^{+1.00}_{-1.10}$ km and $R_{1.4M_\odot}=12.56^{+1.00}_{-1.07}$ km,
indicating the lack of significant variation of the radii from $1.4 M_\odot$ 
to $2.0 M_\odot$. This feature cannot be reproduced by the hadronic matter
due to the softening of equation of state (EoS) by hyperon mixing,
indicating the possible existence of quark phases in neutron-star interiors.
Two models are used for quark phases:
In the quark-hadron transition (QHT) model,
quark deconfinement phase transitions from a hadronic-matter EoS are
taken into account so as to give reasonable mass-radius ($MR$) curves 
by adjusting the quark-quark repulsions and
the density dependence of effective quark mass.
In the quarkyonic model, the degrees of freedom inside the Fermi sea are 
treated as quarks and neutrons exist at the surface of the Fermi sea,
where $MR$ curves are controlled mainly by the thickness of neutron Fermi layer.
The QHT and quarkyonic EoSs can be adjusted so as to reproduce
radii, tidal deformabilities, pressure and central densities inferred 
from the NICER analysis better than the nucleonic matter EoS, 
demonstrating the clear impacts of quark phases.
Then, the maximum mass for the quakyonic-matter EoS is considerably 
larger than that for the QHT-matter EoS.
\end{abstract}

%\pacs{21.30.Cb, 21.45.Ff, 21.65.Cd, 21.80.+a, 25.70.-z, 26.60.Kp}

\maketitle

\parindent 15 pt

\section{Introduction}

In studies of neutron stars (NS), the fundamental role is played by
the equation of state (EoS) for neutron star matter.
The massive neutron stars with masses over $2M_{\odot}$ have been reliably 
established by the observations of NSs J1614$-$2230 \cite{Demorest10}, 
J0348+0432 \cite{Antoniadis13}, J0740+6620 \cite{Cromartie2020} and
J0952-0607 \cite{Romani2022}.
The radius information of NSs have been obtained for the massive
NS PSR J0740+6620 with $2M_{\odot}$ and $1.4M_{\odot}$ NSs, 
shown as $R_{2M_{\odot}}$ and $R_{1.4M_{\odot}}$, from 
the analyses for the X-ray data taken by the 
{\it Neutron Star Interior Composition Explorer} (NICER)
and the X-ray Multi-Mirror (XMM-Newton) observatory.
The analysis of Miller et al. gives $R_{2.08M_{\odot}}=12.35\pm0.75$ km
and $R_{1.4M_{\odot}}=12.45\pm0.65$ km \cite{Miller2021}.
The analysis of Riley gives $R_{2.08M_{\odot}}=12.39^{+1.30}_{-0.98}$ km
and $R_{1.4M_{\odot}}=12.33^{+0.76}_{-0.81}$ km \cite{Riley2021}.
Legred et al. investigate these measurement's implications for the EoSs,
employing a nonparametric EoS model based on Gaussian processes and
combining information from other X-ray and gravitational wave observations
\cite{Legred2021}.
 
The purpose of this paper is to demonstrate that the radius information
of massive NSs give the important constraints for the neutron-star EoSs.
In our EoS analysis, the following neutron-star radii are adopted 
as critical values to be reproduced: 
\begin{eqnarray}
R_{2.0M_\odot} =  12.41^{+1.00}_{-1.10}\ \ {\rm km} 
\nonumber
\\
R_{1.4M_\odot} =  12.56^{+1.00}_{-1.07}\ \ {\rm km} 
\label{radii} 
\end{eqnarray}
with maximum mass $M_{max}/M_\odot = 2.21^{+0.31}_{-0.21}$, being
given by the analysis by Legred et al.\cite{Legred2021}.
The median values of $R_{2M_\odot}$ and $R_{1.4M_\odot}$ in the
above three references \cite{Miller2021}\cite{Riley2021}\cite{Legred2021}
are only a few hundred meters apart from each other.
We set the fitting accuracy to a few hundred meters in our analysis for 
$R_{2M_\odot}$ and $R_{1.4M_\odot}$. Then, the EoS obtained 
from our analysis are not changed, even if the set of 
$R_{2M_\odot}$ and $R_{1.4M_\odot}$ in \cite{Miller2021} or 
\cite{Riley2021} is used as the criterion instead of Eq.(\ref{radii}) or 
all three sets in \cite{Miller2021}\cite{Riley2021}\cite{Legred2021} are used.
The key feature found commonly in the three sets is the small variation
of radii from $1.4M_\odot$ to $2M_\odot$, namely 
$R_{2M_\odot} \approx R_{1.4M_\odot}$.
The reason why the result in \cite{Legred2021} is used in
our present analysis is because they present the inferred values 
of maximum masses, radii, tidal deformabilities,
pressure and central densities obtained from their analysis.
These quantities can be compared with our corresponding results,
by which the features of our EoSs are revealed in detail.

The hyperon mixing in neutron-star matter brings about 
a remarkable softening of the EoS and a maximum mass is reduced
to a value far less than $2M_{\odot}$.
The EoS softening is caused by changing of high-momentum neutrons
at Fermi surfaces to low-momentum hyperons via strangeness 
non-conserving weak interactions overcoming rest masses of hyperons.
In order to derive EoSs for massive NSs, it is necessary to
solve this ``hyperon puzzle in neutron stars".
There have been proposed possible mechanisms:
(i) more repulsive hyperon-hyperon interactions in 
relativistic mean field (RMF) models driven by vector mesons exchanges
\cite{Bednarek,Weiss,Oertel,Maslov}, 
(ii) repulsive hyperonic three-body forces
\cite{NYT,Vidana11,YFYR14,YFYR16,YTTFYR17,Lonardoni,Logoteta,Cerstung},
(iii) appearance of other hadronic degrees of freedom, such as
$\Delta$ isobars \cite{Drago} or meson condensates 
\cite{Kaplan,Brown,Thorsson,Lee,GleSch},
(iv) existence of quark phases in high-density regions 
\cite{Baldo2006,Ozel,Weissenborn, Klahn, Bonanno,Lastowiecki2012,Shahrbaf1,Shahrbaf2,Otto2020,KBH2022,YYR2022}.
It should be noted that the criterion for NS radii Eq.(\ref{radii}) is 
stricter than the condition of $M_{max}>2M_\odot$ only to solve the "puzzle" 
and the above mechanisms are needed to be re-investigated under this stricter condition.

One of the approaches belonging to (ii) is to assume that 
three-nucleon repulsions (TNR) \cite{APR98} work universally among 
every kind of baryons as three-baryon repulsions (TBR) \cite{NYT}.
In \cite{YFYR14,YFYR16,YTTFYR17}, the multi-pomeron exchange potential 
(MPP) was introduced as a model of universal repulsions among three and 
four baryons on the basis of the extended soft core (ESC) baryon-baryon 
interaction model developed by two of the authors 
(T.R. and Y.Y.) and M.M. Nagels \cite{ESC16I,ESC16II,ESC16III}.
In the case of this special modeling for hyperonic three-body repulsions,
the EoS softening by hyperon mixing is not completely recovered by the 
above universal repulsions, and the maximum masses become not so large 
even if universal many-body repulsions increase. 
As a result, the maximum masses for hyperonic-matter EoS cannot be over 
$2M_{\odot}$, as found in \cite{YFYR14,YFYR16,YTTFYR17}:
It is difficult that criterion Eq.(\ref{radii}) is realized by 
this modeling of hadronic-matter EoSs.
A simple way to avoid the strong softening of EoS by hyperon mixing 
is to assume $\Lambda NN$ repulsions stronger than $NNN$ repulsions 
with neglect of $\Sigma^-$ mixing \cite{Lonardoni}. 

In this paper, we focus on the mechanism (iv).
It is possible to solve the ``hyperon puzzle" by taking account of
quark deconfinement phase transitions from a hadronic-matter EoS to
a sufficiently stiff quark-matter EoS in the neutron-star interiors,
namely by studying hybrid stars having quark matter in their cores,
where repulsive effects in quark phases are needed to result in 
massive stars over $2M_{\odot}$. In the Nambu-Jona-Lasinio (NJL) model, 
for instance, repulsions to stiffen EoSs are given by vector interactions.
Then, it is known well that quark-hadron phase transitions should
be crossover or at most weak first-order, because strong first-order
transitions soften EoSs remarkably in order to obtain stiff EoSs.
In \cite{KBH2022}, they derived the new EoS within the quark-hadron 
crossover (QHC) framework (3-windows model) so as to reproduce
$R_{2.1M_\odot} \approx R_{1.4M_\odot} \approx  12.4 {\rm km}$.
Here, the small variation of radii indicates that
the pressure grows rapidly while changes in energy density are modest,
producing a peak in the speed of sound \cite{KBH2022}. 
In their QHC framework, the EoSs in the quark-hadron mixed region of 
$1.5\rho_0 \sim 3.5\rho_0$, playing a decisive role for the resulting
$MR$ curves,  are given by the interpolating functions phenomenologically.
Then, it is meaningful to study the other modeling for phase transitions
in which the mixed regions are modeled explicitly.
We investigate how this criterion Eq.(\ref{radii}) can be 
realized in the case of using the EoS derived from our quark-hadron transition
(QHT) model for neutron-star matter in the Bruecner-Hartree-Fock (BHF) 
framework \cite{YYR2022}, being different from their 3-windows model.
Here, the quark-matter EoS is derived from the two-body quark-quark ($QQ$) 
potentials, in which all parameters are on the physical backgrounds with 
no room for arbitrarily changing:
They are composed of meson-exchange quark-quark potentials 
derived by unfolding of the baryon-baryon meson-exchanges, and
instanton-exchange, one-gluon-exchange and multi-pomeron exchange potentials.
Then, baryonic matter and quark matter are treated in the common BHF framework, 
where quark-hadron transitions are treated on the basis of the Maxwell condition. 
In this paper, it is shown that the criterion Eq.(\ref{radii}) can be 
realized by our QHT model for neutron-star matter, as well as the QHC model
\cite{KBH2022}, by adjusting the $QQ$ repulsion to be strong enough 
and the quark-hadron transition density to be about $2\rho_0$. 

In our QHT model the BHF framework is used for deriving the 
quark-matter EoS, which is not popular. Our treatments for 
quark-hadron phase transitions is the same as that in \cite{Shahrbaf2}
where the NJL model is adopted for quark matter under the mean field
approximation. In spite of the difference between quark-matter models,
their obtained $MR$ curves are similar to ours in \cite{YYR2022}.
Therefore, it is considered that the same conclusions can be derived
also by using their QHT model instead of ours.

Another type of quark phase in neutron-star interiors is given by
the quarkyonic matter
\cite{MP2007,HMP2008,MR2019,HMLCP2019,DHJ2020,ZL2020,MHPC2021,Cao2022}, 
where the degrees of freedom inside the Fermi sea are treated as quarks
and nucleons exist at the surface of the Fermi sea.
The transition from hadronic-matter phase to the quarkyonic-matter phase
is considered to be in second-order.
In the quarkyonic matter, the existence of free quarks inside
the Fermi sea gives nucleons extra kinetic energy by pushing them
to higher momenta, leading to increasing pressure.
This mechanism to realize the criterion Eq.(\ref{radii})
is completely different from the QHT matter in which the 
essential roles for EoS stiffening are played by the $QQ$ repulsions. 
Then, it is valuable to study the characteristic differences
between neutron-star mass-radius ($MR$) curves obtained from the 
QHT-matter EoS and quarkyonic-matter EoS. 

This paper is organized as follows:
In Sect.II, the hadronic-matter EoS (II-A), 
the quark-matter EoS (II-B) and the quarkyonic-matter EoS (II-C)
are formulated on the basis of our previous works, where 
the BHF frameworks with our $QQ$ potentials are adopted both 
for baryonic matter and quark (quarkyonic) matter.
Transitions from hadron phases to quark matter (quakyonic) phases
are explained.
In Sect.III-A, the calculated results are shown for pressures,
energy densities and sound velocities.
In III-B, the $MR$ curves of hybrid stars are obtained 
by solving the Tolmann-Oppenheimer-Volkoff (TOV) equation.
In III-C, the obtained values of maximum masses, radii, tidal deformabilities,
pressure and central densities are compared with those inferred from
the NICER-data analysis.
The conclusion of this paper is given in Sect.IV.

\section{Models of neutron-star matter}

\subsection{hadronic matter}

The hadronic matter is defined as $\beta$-stable hyperonic nuclear matter
including leptons, composed of $n$, $p^+$, $\Lambda$, $\Sigma^-$, $e^-$, $\mu^-$. 
We recapitulate here the hadronic-matter EoS.
In the BHF framework, the EoS is derived with use of the 
ESC baryon-baryon ($B\!B$) interaction model \cite{YFYR14,YFYR16,YTTFYR17}. 

As is well known, 
the nuclear-matter EoS is stiff enough to assure neutron-star masses
over $2M_{\odot}$, if the strong three-nucleon repulsion (TNR) is taken into account.
However, there appears a remarkable softening of EoS by inclusion of exotic degrees 
of freedom such as hyperon mixing. 
One of the ideas to avoid this ``hyperon puzzle" is to assume that 
the many-body repulsions work universally for every kind of baryons \cite{NYT}.
In \cite{YFYR14,YFYR16,YTTFYR17}, 
the multi-pomeron exchange potential MPP was introduced as a model of universal 
repulsions among three and four baryons. This was inspired by the multi-reggeon model
to describe CERN-ISR pp-data \cite{KM74}.
The ESC work is mentioned in \cite{ESC16I,ESC16II,ESC16III}. 

In \cite{YTTFYR17} they proposed three versions of MPP (MPa, MPa$^+$, MPb),
where MPa and MPa$^+$ (MPb) include the three- and four-body (only three-body) repulsions.
Their strengths are determined by analyzing the nucleus-nucleus scattering 
using the G-matrix folding model under the conditions that the saturation
parameters are reproduced reasonably.
The EoSs for MPa and MPa$^+$ are stiffer than that for MPb, and
maximum masses and radii of neutron stars obtained from MPa, MPa$^+$
are larger than those from MPb. 
The important criterion for repulsive parts is the resulting neutron-star radii $R$ 
for masses of $1.4M_\odot$: In the case of using MPb, we obtain  
$R_{1.4M_\odot} \approx 12.4$ km similar to the value in the criterion Eq.(\ref{radii}).
On the other hand, we have $R_{1.4M_\odot} \approx 13.3$ (13.6) km
in the case of MPa (MPa$^+$).
In this paper, we adopt MPb as three-baryon repulsion:
Our nuclear interactions are composed of 
two-body part $V_{BB}$ and three-body part $V_{BBB}$, where $V_{BB}$ and 
$V_{BBB}$ are given by ESC and MPb, respectively.
It is worthwhile to say that the three-nucleon repulsion in MPb is stronger 
than the corresponding one (UIX) in the standard model by APR \cite{APR98} 
giving rise to $R_{1.4M_\odot} \approx 11.6$ km \cite{Togashi1}.

$BB$ G-matrix interactions ${\cal G}_{BB}$ are derived from
$BB$ bare interactions $V_{BB}$ or $V_{BB}+V_{BBB}$ \cite{YFYR14}. 
They are given for each $(BB',T,S,P)$ state, $T$, $S$ and $P$
being isospin, spin and parity in a two-body state, respectively,
and represented as ${\cal G}_{BB'}^{TSP}$.
The G-matrix interactions derived from $V_{BB}$ and
$V_{BB} +V_{BBB}$ are called B1 and B2, respectively.
In the quarkyonic model, we need only the 
neutron-neutron sectors, ${\cal G}_{nn}^{SP}$.

A single baryon potential is given by
\begin{eqnarray}
U_B(k)&=&\sum_{B'=n,p,\Lambda,\Sigma^-} U_{B}^{(B')}(k) 
\nonumber\\
 &=& \sum_{B'=n,p,\Lambda,\Sigma^-}  \sum_{k'<k_F^{(B')}} 
 \langle kk'|{\cal G}_{BB'}|kk'\rangle
\nonumber\\
\end{eqnarray}
with $B=n, p,\Lambda,\Sigma^-$. 	
Here, $\langle kk'|{\cal G}_{BB'}|kk'\rangle$ is 
a $BB'$ G-matrix element in momentum space,
being derived from $V_{BB}$ or ($V_{BB}$+$V_{BBB}$),
and $k_F^{(B)}$ is the Fermi momentum of baryon $B$.
In this expression, spin and isospin quantum numbers are implicit.

The baryon energy density is given by
\begin{eqnarray}
\varepsilon_B&=& \tau_B + \upsilon_B 
\nonumber \\
  &=& g_s \int_{0}^{k_F^{(B)}} \frac{d^3k}{(2\pi)^3}
\left\{\sqrt{\hbar^2 k^2+M_B^2}+\frac 12 U_B(k)\right\} \ ,
\nonumber \\
\label{edenn} 
\end{eqnarray}
where $\tau_B$ and $\upsilon_B$ are kinetic and
potential parts of the energy density.

In $\beta$-stable hadronic matter composed of
$n$, $p$, $e^-$, $\mu^-$, $\Lambda$ and $\Sigma^-$,
equilibrium conditions are given as

\medskip
\noindent
(1) chemical equilibrium conditions,
\begin{eqnarray}
\label{eq:c1}
&& \mu_n = \mu_p+\mu_e \\
&& \mu_\mu = \mu_e \\
&& \mu_\Lambda = \mu_n \\
&& \mu_{\Sigma^-} =\mu_n + \mu_e 
%&& \mu_{\Xi^-} =\mu_n + \mu_e
\label{eq:c2}
\end{eqnarray}
\noindent
(2) charge neutrality,
\begin{eqnarray}
\rho_p = \rho_e +\rho_\mu+\rho_{\Sigma^-}
\end{eqnarray}
\noindent
(3) baryon number conservation,
\begin{eqnarray}
\rho = \rho_n +\rho_p +\rho_\Lambda+\rho_{\Sigma^-} \ .
\label{eq:c3}
\end{eqnarray}
\noindent

Expressions for $\beta$-stable nucleonic matter composed of 
$n$, $p$, $e^-$ and $\mu^-$ are obtained by omitting hyperon sectors 
from the above expressions for $\beta$-stable baryonic matter.

\subsection{Quark-Hadron transition model}

In our treatment of quark matter, the BHF framework is adopted 
on the basis of two-body $QQ$ potentials \cite{YYR2022}.
Here, correlations induced by bare $QQ$ potentials are renormalized into 
coordinate-space G-matrix interactions, being considered as effective 
$QQ$ interactions used in quark-matter calculations. 

Our bare $QQ$ interaction is given by
\begin{eqnarray}
V_{QQ} &=& V_{EME}+V_{INS}+ V_{OGE}+V_{MPP} 
\end{eqnarray}
where $V_{EME}$, $V_{INS}$, $V_{OGE}$ and $V_{MPP}$ are the extended 
meson-exchange potential, the instanton-exchange potential, the 
one-gluon exchange potential and the multi-pomeron exchange potential, 
respectively. Parameters in our $QQ$ potential are chosen 
so as to be consistent with physical observables.
The $V_{EME}$ $QQ$ potential is derived from the ESC $BB$ potential 
so that the $QQM$ couplings are related to the $BBM$ couplings
through folding procedures with Gaussian baryonic quark wave functions.
In the construction of the relation between $BBM$ and $QQM$ couplings,
the requirement that the coefficients of the $1/M^2$ expansion should match 
is based on Lorentz invariance, which fixes the QQM couplings and also 
determines the (few) extra vertices at the quark level \cite{ESC16I}.
Then, the $V_{EME}$ $QQ$ potential is basically of the same functional 
expression as the ESC $BB$ potential. 
Strongly repulsive components in ESC $BB$ potentials
are described mainly by vector-meson and pomeron exchanges between baryons.
This feature persists in the $V_{EME}$ $QQ$ potential,
which includes the strongly repulsive components originated from
vector-meson and pomeron exchanges between quarks.
Similarly the multi-pomeron exchange potentials among quarks, $V_{MPP}$, 
are derived from the corresponding ones among baryons, giving
repulsive contributions.
Contributions from $V_{INS}$ and $V_{OGE}$ in average are attractive 
and repulsive, respectively. 
The strength of $V_{OGE}$ is determined by the quark-gluon coupling 
constant $\alpha_S$. 
In \cite{YYR2022} $\alpha_S$ is chosen as 0.25, that is $V_{OGE}(\alpha_S=0.25)$,
and the three sets are defined as follows:
Q0 : $V_{EME}$, Q1 : $V_{EME}+V_{INS}+V_{OGE}(\alpha_S=0.25)$
Q2 : $V_{EME}+V_{MPP}+V_{INS}+V_{OGE}(\alpha_S=0.25)$.

In our QHT model for neutron-star matter, quark-hadron phase transitions occur 
at crossing points of hadron pressure $P_H(\mu)$ and quark pressure $P_Q(\mu)$ 
being a function of chemical potential $\mu$. 
Positions of crossing points, giving quark-hadron transition densities, 
are controlled by parameters $\rho_c$ and $\gamma$ included 
in our density-dependent quark mass 
\begin{eqnarray}
M_Q^*(\rho_Q) = \frac{M_0}{1+\exp [\gamma (\rho_Q-\rho_c)]} +m_0 +C
\label{mstar}
\end{eqnarray}
with $C=M_0-M_0/[1+\exp (-\gamma \rho_c)]$ assuring $M_Q^*(0) = M_0+m_0$,
where $\rho_Q$ is number density of quark matter, and $M_0$ and $m_0$ 
are taken as 300 (360) MeV and 5 (140) MeV for $u$ and $d$ ($s$) quarks.
Here, the effective quark mass $M_Q^*(\rho_Q)$ should be used together
with $B(\rho_Q)=M_Q^*(0) -M_Q^*(\rho_Q)+B_0$, meaning the energy-density
difference between the perturbative vacuum and the true vacuum.
A constant term $B_0$ is added for fine tuning of an onset density.
In \cite{YYR2022}, the values of ($\rho_c$, $\gamma$) without $B_0$ 
are given for each set of Q0, Q1 and Q2.

Let us focus on the typical result for Q2+H1 in \cite{YYR2022}.
The $QQ$ interaction Q2 is the most repulsive among Q0, Q1 and Q2.
The $BB$ interaction H1 consists of ESC and MPb, and results in 
the reasonable value of $R_{1.4M_\odot}$.
In this case of Q2+H1, we obtain the maximum mass of 
$2.25M_\odot$ and the reasonable value of $R_{1.4M_\odot}=12.5$ km, 
in which the quark-hadron transition occurs at density of 3.5$\rho_0$.
Then, we have $R_{2.0M_\odot}=12.0$ km, being rather        
smaller than 12.4 km in the criterion Eq.(\ref{radii}).
In order to reproduce a larger value of $R_{2.0M_\odot} \approx 12.4$ km, 
we make $V_{OGE}$ more repulsive by taking larger values of 
$\alpha_S=$ 0.36 and 0.49. 
It is not suitable for such a purpose to strengthen the $V_{MPP}$ repulsion,
because $V_{MPP}$ is essentially of three-body interaction
and the contributions in low-density region are small.
On the other hand, $V_{OGE}$ is of two-body interaction, and 
its repulsive contributions are not small even in low density region,
being important for a large value of $R_{2.0M_\odot}$.
Another condition to make $R_{2.0M_\odot}$ larger is
to lower quark-hadron transition densities by adjusting
the parameters ($\rho_c$,$\gamma$,$B_0$) included in the 
density-dependent quark mass Eq.(\ref{mstar}).

We define newly the following three sets with the fixed value of $\gamma$=1.2
\medskip
\par
\noindent
Q2 : $V_{EME}+V_{MPP}+V_{INS}+V_{OGE}(\alpha_S=0.25)$ \ 
\par
\noindent
\qquad  \quad with $\rho_c=6.9\rho_0$ and $B_0=$8.5 

\noindent
Q3 : $V_{EME}+V_{MPP}+V_{INS}+V_{OGE}(\alpha_S=0.36)$ \ 
\par
\noindent
\qquad  \quad with $\rho_c=6.9\rho_0$ and $B_0=$7.5 

\noindent
Q4 : $V_{EME}+V_{MPP}+V_{INS}+V_{OGE}(\alpha_S=0.69)$ \ 
\par
\noindent
\qquad  \quad with $\rho_c=7.5\rho_0$ and $B_0=$10.0
\par
\noindent
where the values of $\rho_c$ and $B_0$ for each set are chosen 
so as to give quark-hadron transition densities of $\sim 2 \rho_0$.

G-matrix interactions ${\cal G}_{qq'}$ with $q,q'=u,d,s$ 
are derived from the above bare $QQ$ interactions.
They are given for each $(qq',T,S,P)$ state, $T$, $S$ and $P$
being isospin, spin and parity in a two-body state, respectively,
and represented as ${\cal G}_{qq'}^{TSP}$.
Hereafter, Q2, Q3 and Q4 mean the naming of corresponding 
$QQ$ G-matrix interactions, not only of bare $QQ$ interactions. 
The $QQ$ G-matrix interactions are used also in 
the quarkyonic matter calculations.

\medskip

A single quark potential is given by
\begin{eqnarray}
U_q(k)&=&\sum_{q'=u,d,s} U_{q}^{(q')}(k) 
 = \sum_{q'=u,d,s} \sum_{k'<k_F^{q'}} \langle kk'|{\cal G}_{qq'}|kk'\rangle
\nonumber\\
\end{eqnarray}
with $q=u,d,s$, where $k_F^{q}$ is the Fermi momentum of quark $q$.
Spin and isospin quantum numbers are implicit.

The quark energy density is given by
\begin{eqnarray}
\varepsilon_q&=&
 g_s N_c\sum_{q=u,d,s} \int_0^{k_{Fq}} \frac{d^3k}{(2\pi)^3}
\nonumber \\
&& \left\{\sqrt{\hbar^2 k^2+M_q^2}+\frac 12 U_q(k) \right\} \ .
\nonumber \\
\label{edenq} 
\end{eqnarray}
Fermion spin and quark color degeneracies give rise to $g_s=2$ and $N_c=3$.

\begin{figure}[ht]
\begin{center}
\includegraphics*[width=8.6cm]{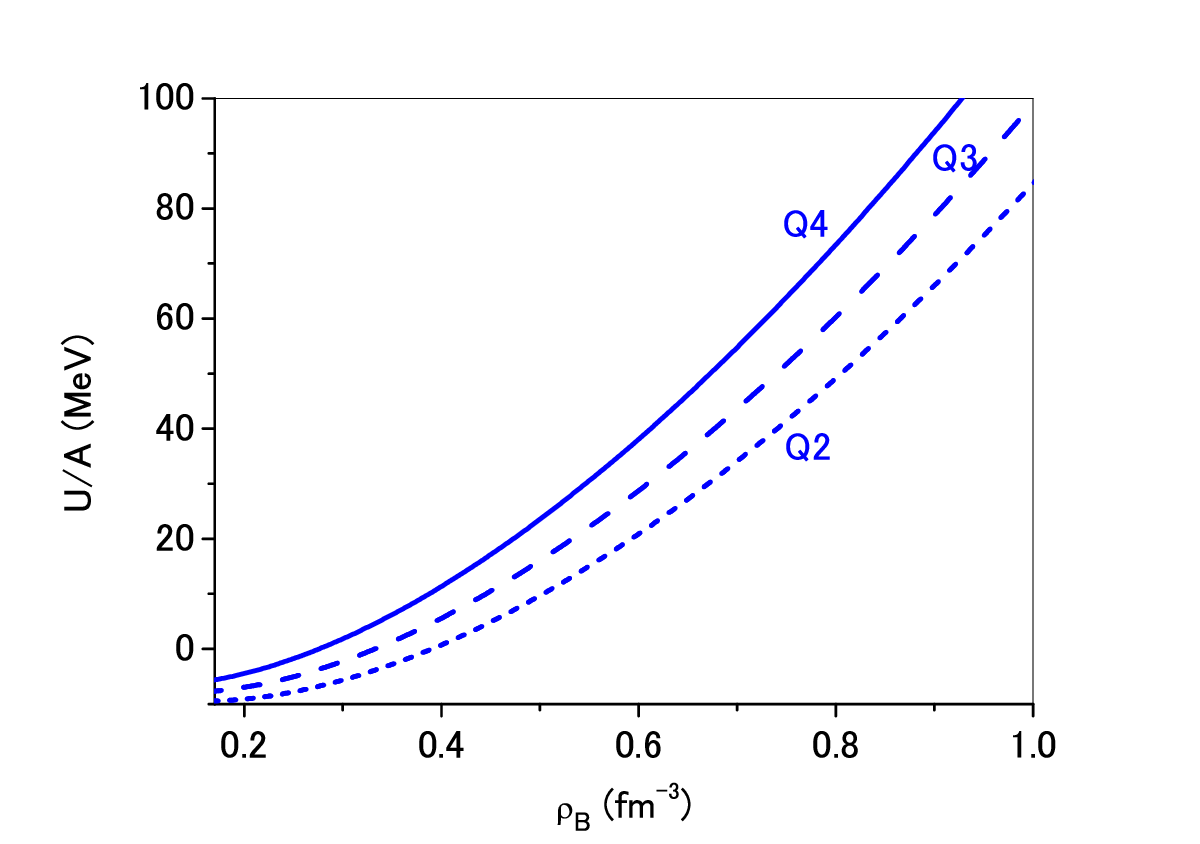}
\caption{(Color online) Potential energies per particle $U/A$ as 
a function of the baryon number density $\rho_B$ in the case of 
$\rho_u=\rho_d=\rho_s$. The short-dashed, long-dashed and solid 
curves are obtained by using Q2, Q3 and Q4, respectively.
}
\label{Ucon}
\end{center}
\end{figure}

In order to demonstrate the features of our $QQ$ interactions
(Q2,Q3,Q4), we show the potential energy per particle $U/A$ 
as a function of the baryon number density $\rho_B=\frac13 \rho_Q$ 
in the case of taking $\rho_u=\rho_d=\rho_s$.
In Fig.\ref{Ucon}, the short-dashed, long-dashed and solid curves
are obtained by using Q2, Q3 and Q4, respectively.
The repulsions are found to be strong in the order of Q4, Q3, Q2.
This difference of repulsions among Q4, Q3 and Q2 comes from
the different values of $\alpha_S$ included in $V_{OGE}$.
In the figure, it should be noted that the difference is 
not small even in the low-density region.

In the EoS of $\beta$-stable quark matter 
composed of $u$, $d$, $s$, $e^-$,
the equilibrium conditions are given as

\medskip
\noindent
(1) chemical equilibrium conditions,
\begin{eqnarray}
&& \mu_d = \mu_s = \mu_u+\mu_e 
\label{eq:c1}
\end{eqnarray}
\noindent
(2) charge neutrality,
\begin{eqnarray}  
     0 = \frac13 (2\rho_u -\rho_d -\rho_s) -\rho_e 
\label{eq:c2}
\end{eqnarray}
\noindent
(3) baryon number conservation,
\begin{eqnarray}
\rho_B = \frac13 (\rho_u +\rho_d +\rho_s)= \frac13 \rho_Q \ .
\label{eq:c3}
\end{eqnarray}

In order to construct the hybrid EoS including a transition from 
hadronic phase to quark phase, we use the replacement interpolation 
method  \cite{Shahrbaf2} \cite{YYR2022},
being a simple modification of the Maxwell and the Glendenning (Gibbs) 
constructions \cite{Glendenning}. 
The EoSs of hadronic and quark phases and that of mixed phase are
described with the relations between pressures and chemical potentials
$P_H(\mu)$, $P_Q(\mu)$ and $P_M(\mu)$, respectively.
The critical chemical potential $\mu_c$ for the transition
from the hadronic phase to the quark phase is 
obtained from the Maxwell condition 
\begin{eqnarray}
P_Q(\mu_c)=P_H(\mu_c)=P_c  \ .
\end{eqnarray}
The pressure of the mixed phase is represented by a polynomial ansatz.
The matching densities $\rho_H$ and $\rho_Q$ are obtained
with use of $\rho(\mu)=dP(\mu)/d\mu$.

\subsection{quarkyonic matter}
In the BHF framework,
we derive the EoS of quarkyonic matter composed of neutrons 
and quarks with flavor $q=u,d$ in the simplest form
by McLerran and Reddy \cite{MR2019}.
In the chargeless 2-flavor quarkyonic matter, strongly interacting
quarks near the Fermi sea form interacting neutrons, and
the remaining d and u quarks fill the lowest momenta up to
$k_{Fu}$ and $k_{Fd}$, respectively.  The quark mass is taken 
to be $M_q=M_n/3$ constantly, $M_n$ being the neutron mass.
In calculations of quarkyonic matter, 
we use B1 ($V_{nn}$) and B2 ($V_{nn}$+$V_{nnn}$) for nuclear interactions,
and Q0 for $QQ$ interactions for simplicity. 

The total baryon number density is given by
\begin{eqnarray}
\rho_B &=& \rho_n+\frac{N_c}{3} (\rho_u+\rho_d)
\nonumber \\
&=& \frac{g_s}{6\pi^2} \left[k_{Fn}^3-k_{0n}^3 +\frac{N_c}{3}
(k_{Fu}^3+k_{Fd}^3) \right] \ ,
\end{eqnarray}
where $k_{Fn}$, $k_{Fu}$ and $k_{Fd}$ are the Fermi momenta of
neutrons and u and d quarks, respectively. Fermion spin and 
quark color degeneracies give rise to $g_s=2$ and $N_c=3$.
Neutrons are restricted near the Fermi surface by $k_{0n}$,
being assumed as
\begin{eqnarray}
&& k_{0n} = k_{Fn} -\Delta_{qyc}
\nonumber \\
&& \Delta_{qyc} = \frac{\Lambda^3}{\hbar c^3 k_{Fn}^2} 
      +\kappa \frac{\Lambda}{N_c^2 \hbar c } \ ,
\end{eqnarray}
where $\Delta_{qyc}$ for the thickness of Fermi layer
includes the two parameters $\Lambda$ and $\kappa$.
In this work, we take the fixed value of $\kappa=0.3$.

Then, $k_{Fd}$ and $k_{Fu}$ are related to $k_{0n}$
by $k_{Fd}=\frac{1}{N_c}k_{0n}$ and $k_{Fu}=2^{-1/3} k_{Fd}$.

A single neutron potential is given by
\begin{eqnarray}
U_n(k)&=& \sum_{k_{0n}<k'<k_{Fn}} \langle kk'|{\cal G}_{nn}|kk'\rangle 
\label{Unn} 
\end{eqnarray}
with $nn$ G-matrix interactions ${\cal G}_{nn}$.

The neutron energy density is given by
\begin{eqnarray}
\varepsilon_n&=& \tau_n + \upsilon_n 
\nonumber \\
  &=& g_s \int_{k_{0n}}^{k_{Fn}} \frac{d^3k}{(2\pi)^3}
\left\{\sqrt{\hbar^2 k^2+M_n^2}+\frac 12 U_n(k)\right\} \ .
\nonumber \\
\label{edenn} 
\end{eqnarray}

Additionally, another form of the neutron potential 
energy density is defined as
\begin{eqnarray}
%&& V_n=
&& {\bar \upsilon_n}= 
 g_s \int_{0}^{k_{n}} \frac{d^3k}{(2\pi)^3}
\left\{\frac 12 U_n(k)\right\} \ ,
\label{vden} 
\end{eqnarray}
which is used in \cite{MR2019} instead of $\upsilon_n$.

Single quark potentials for $q=u, d$ are given by
\begin{eqnarray}
U_q(k) &=& \sum_{q'=u,d} U_{q}^{(q')}(k) 
\nonumber \\
&& = \sum_{q'=u,d} \sum_{k'<k_{Fq}} \langle kk'|{\cal G}_{qq'}|kk'\rangle
\label{Uqq} 
\\
U_q^{(n)}(k) 
 &=& \sum_{k_{0n}<k'<k_{Fn}} \langle kk'|{\cal G}_{qn}|kk'\rangle
\label{Uqn} 
\end{eqnarray}
with G-matrix interactions ${\cal G}_{qq'}$ and ${\cal G}_{qn}$.
Here, ${\cal G}_{qn}$ is the quark-neutron ($Qn$) interactions:
We assume the simple model in which the potentials 
${\cal G}_{qq'}$ are folded into the potentials 
${\cal G}_{qn}$ with Gaussian baryonic quark wave functions.
In Eqs.(\ref{Unn})(\ref{Uqq})(\ref{Uqn})
spin quantum numbers are implicit.

The quark energy density is given by
\begin{eqnarray}
\varepsilon_q&=&
 g_s N_c\sum_{q=u,d} \int_0^{k_{Fq}} \frac{d^3k}{(2\pi)^3}
\nonumber \\
&& \left\{\sqrt{\hbar^2 k^2+M_q^2}+\frac 12 U_q(k)+U_{qn}(k)\right\} \ ,
\nonumber \\
\label{edenq} 
\end{eqnarray}
where values of $k_{Fq}$ are determined by
\begin{eqnarray}
N_c k_{Fq} = k_{0n} \ .
\end{eqnarray}
Thus, our total energy density is given by
\begin{eqnarray}
\varepsilon=\varepsilon_n+\varepsilon_d+\varepsilon_u \ .
\end{eqnarray}
The chemical potential $\mu_i$ ($i=n,d,u$) and pressure $P$
are expressed as
\begin{eqnarray}
&&\mu_i = \frac{\partial \varepsilon_i}{\partial n_i} \ , 
\label{chem} \\
&& P = \sum_{i=n,d,u} \mu_i n_i -\varepsilon \ ,
\label{press}
\end{eqnarray}
where $\frac{\partial \varepsilon_i}{\partial n_i}=
\frac{\partial \varepsilon_i}{\partial n_B}
\frac{\partial n_B}{\partial n_i}$ .

\medskip

In our model, the phase transition from $\beta$-stable nucleonic matter to 
the quarkyonic matter occurs in second-order, resulting in the hybrid EoS 
including hadronic and quarkyonic EoSs. Then, the transition densities 
are controlled mainly by the parameter $\Lambda$:
In this work, we choose the three values of $\Lambda$=380, 350 and 320 MeV 
with the fixed value of $\kappa=0.3$. The transition densities for these values
are $0.28 \sim 0.38$ fm$^{-3}$ ($0.28 \sim 0.36$ fm$^{-3}$)
in the case of using B1 (B2) for nuclear interactions.
Hereafter, when a value of $\Lambda$=380 MeV is used, for instance, 
it is denoted as $\Lambda 380$.

\section{Results and discussion}

\subsection{EoS}

\begin{figure}[ht]
\begin{center}
\includegraphics*[width=8.6cm]{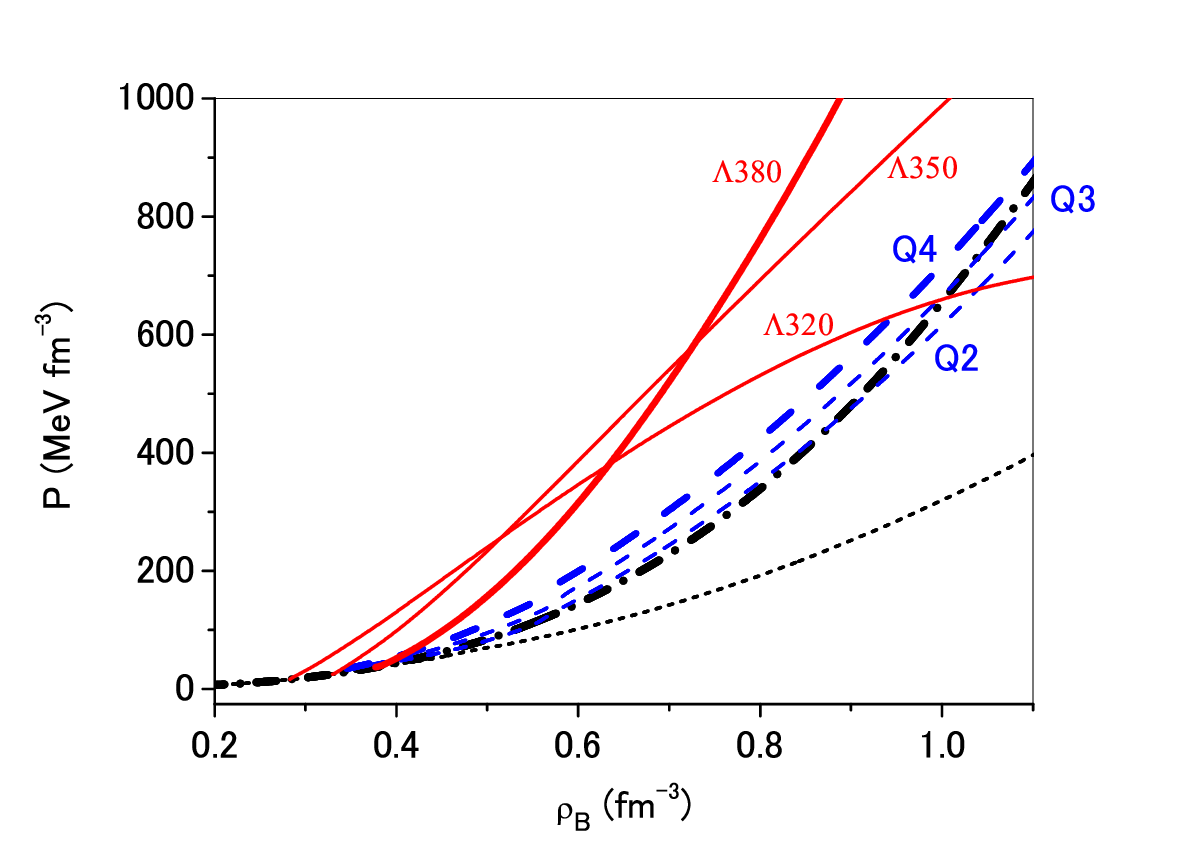}
\caption{(Color online) Pressures $P$ as a function of baryonic number density $\rho_B$.
The dot-dashed (dotted) curve is for $\beta$-stable nucleonic (hadronic) matter. 
Upper thin (thick) solid curves are pressures in the quarkyonic matter for 
$\Lambda 350$ and $\Lambda 320$ ($\Lambda 380$) with B1.
Lower thin (thick) short-dashed curves are for the QHT matter with Q2 and Q3 (Q4).}
\label{Pden1} 
\end{center}
\end{figure}

\begin{figure}[ht]
\begin{center}
\includegraphics*[width=8.6cm]{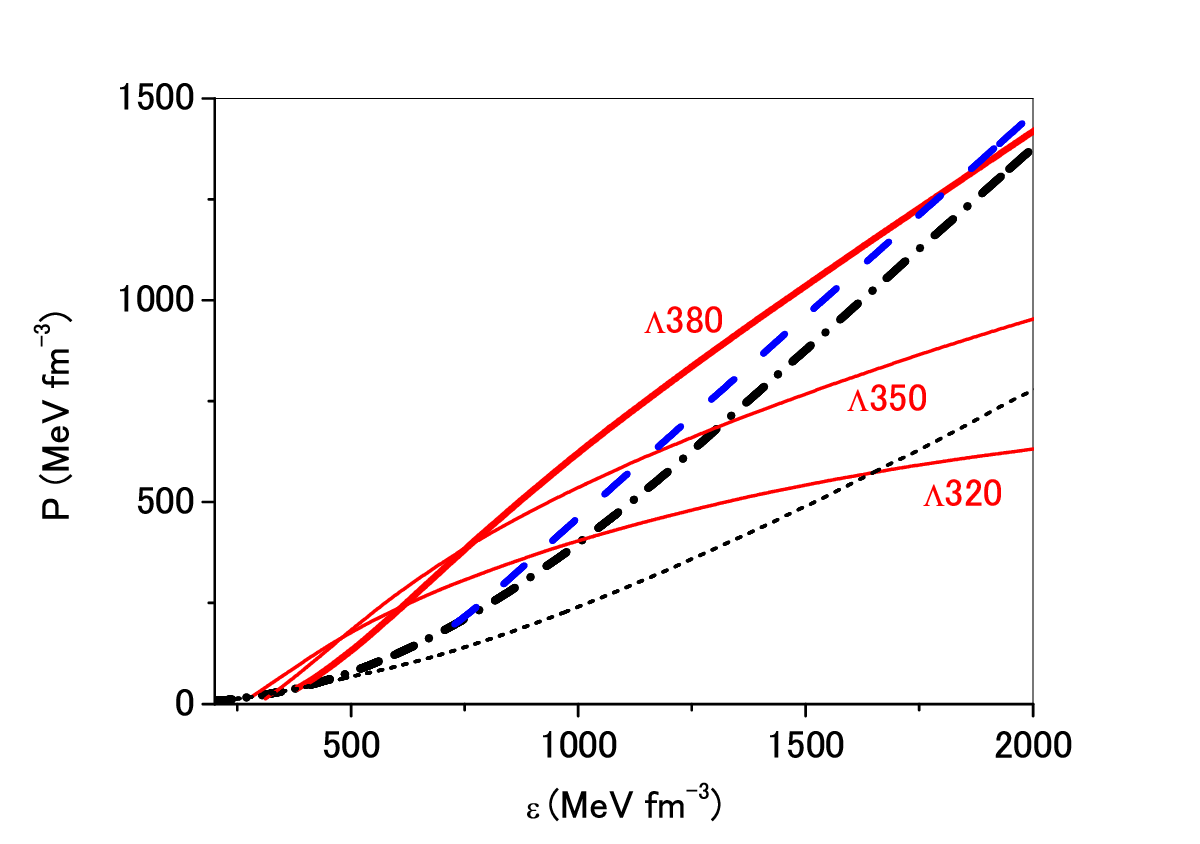}
\caption{(Color online) Pressures $P$ as a function of the energy density $\varepsilon$.
The dot-dashed (dotted) curves are for $\beta$-stable nucleonic (hadronic) matter.
Thin (thick) solid curves show pressures in quarkyonic phases for
$\Lambda 350$ and $\Lambda 320$ ($\Lambda 380$) with B1.
The short-dashed curve is for the QHT model with Q4.}
\label{eden}
\end{center}
\end{figure}

%\begin{figure*}[ht]
\begin{figure}[ht]
\begin{center}
\includegraphics*[width=8.6cm]{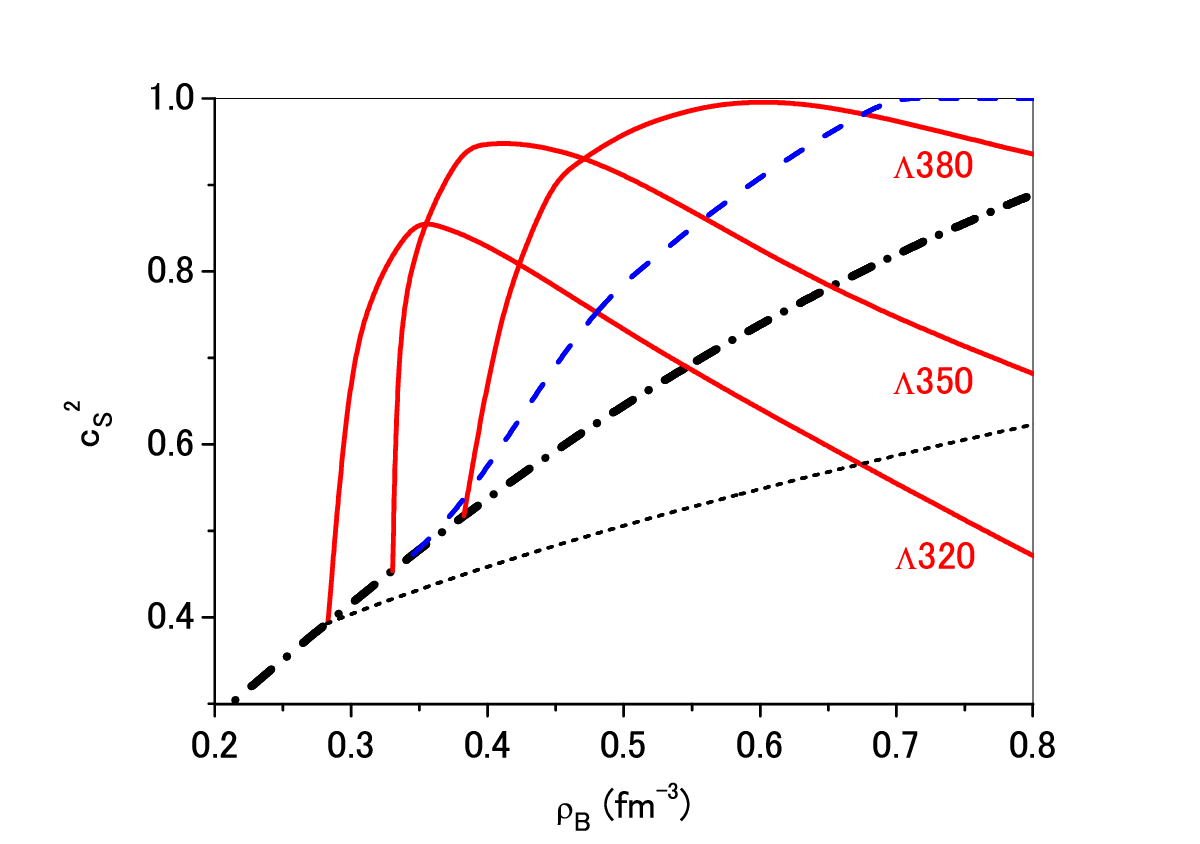}
\caption{(Color online) The square of the sound speed $c_s^2$ in units of $c^2$ 
as a function of baryonic number density $\rho_B$. The dot-dashed (dotted) curve 
is that in $\beta$-stable nucleonic (hadronic) matter.
Solid curves are pressures in quarkyonic matter for $\Lambda 380$, 
$\Lambda 350$ and $\Lambda 320$ with B1.
The dashed curve is for the QHT matter with Q4.}
\label{sound}
\end{center}
\end{figure}
%\end{figure*}

\begin{figure*}[ht]
\begin{center}
\includegraphics*[width=6in,height=3in]{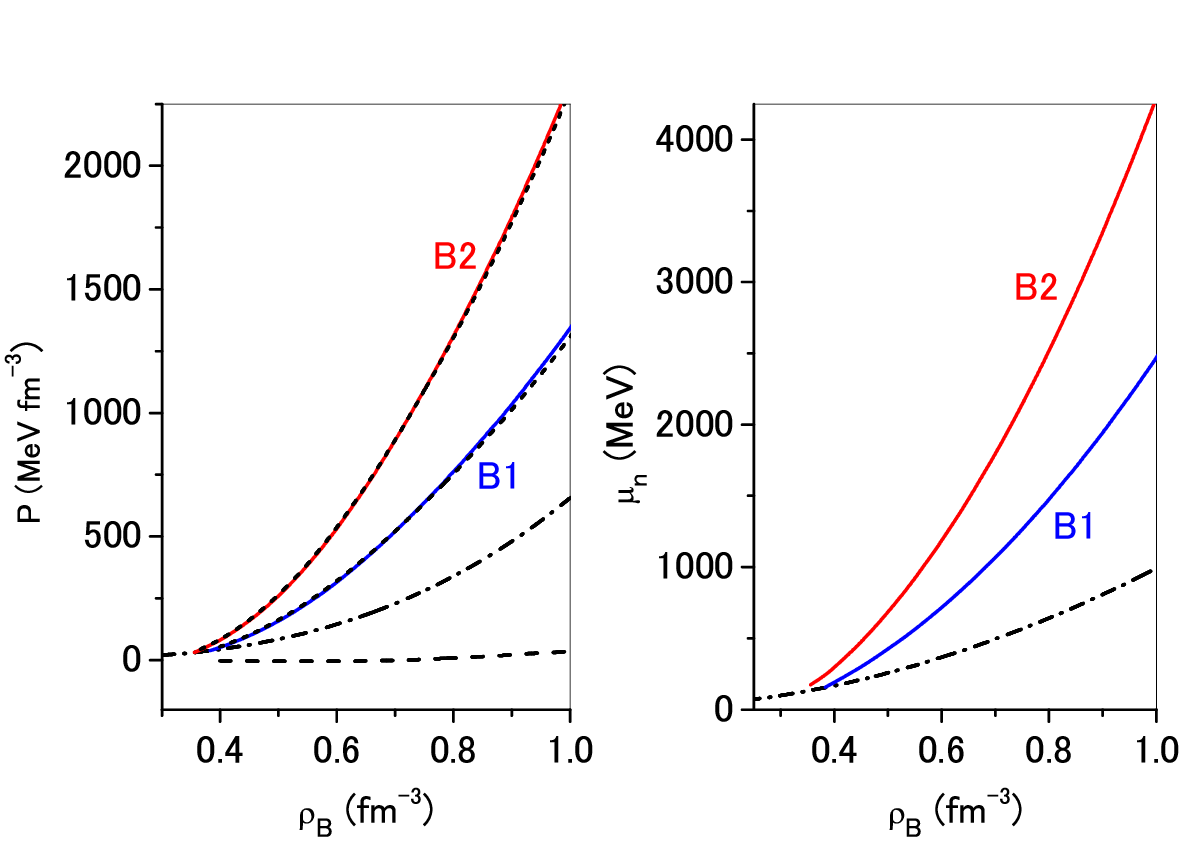}
\caption{(Color online) In the left panel, solid curves are pressures $P$ in quarkyonic 
phases for as a function of baryonic number density $\rho_B$ for $\Lambda 380$
in the cases of using B1 and B2, and short-dashed (dashed) curves are
partial pressures of neutrons (quarks) in respective cases.
The dot-dashed curve is for $\beta$-stable nucleonic matter.
In the right panel, solid (dot-dashed) curves are neutron chemical potentials $\mu_n$
in quarkyonic ($\beta$-stable nucleonic) phases as a function of $\rho_B$
for $\Lambda 380$ in the cases of using B1 and B2. The dot-dashed curve gives 
neutron chemical potential in $\beta$-stable nucleonic matter.}
\label{Pden2}
\end{center}
\end{figure*}

In Fig.\ref{Pden1}, pressures $P$ are drawn as a function of
baryonic number density $\rho_B$. The dot-dashed curve is for 
the $\beta$-stable nucleonic-matter EoS, and the dotted one is for 
the $\beta$-stable hadronic-matter EoS with hyperon mixing.
The latter is substantially below the former, demonstrating the EoS 
softening by hyperon mixing.
Thin (thick) solid curves in the upper side are pressures in the quarkyonic
matter for $\Lambda 350$ and $\Lambda 320$ ($\Lambda 380$) with use of 
B1 for nuclear interactions.
At the crossing points with the dot-dashed curve in the low-density side, 
there occur second-order transitions from $\beta$-stable nucleonic 
to quarkyonic phases:
The transition densities $\rho_t$ are 0.38, 0.33, 0.28 fm$^{-3}$
(2.2$\rho_0$, 1.9$\rho_0$, 1.6$\rho_0$) in the cases of $\Lambda 380$, 
$\Lambda 350$ and $\Lambda 320$, respectively.
Thin (thick) short-dashed curves are for the QHT models with Q2 and Q3 (Q4).
It should be noted that pressures in the quarkyonic matter increase
more rapidly with density than those in the QHT matter.
As discussed later, the rapid growth of pressure with density in the range
of $2\rho_0\sim4\rho_0$ is an important feature of the quarkyonic model.
This rapid increase of pressure at onset of the quarkyonic phase
influences significantly on neutron-star $MR$ curves.

In Fig.\ref{eden}, pressures $P$ are drawn as a function of the 
energy density $\varepsilon$, which are related closely to 
neutron-star $MR$ curves. The dot-dashed (dotted) curve
shows pressures in $\beta$-stable nucleonic (hadronic) matter.
Thin (thick) solid curves show pressures in quarkyonic matter for
$\Lambda 350$ and $\Lambda 320$ ($\Lambda 380$) with B1.
The short-dashed curve is for the QHT matter with Q4.
Though the curves for Q4 and $\Lambda 380$ are rather similar 
to each other in comparison with the corresponding curves
in Fig.\ref{Pden1}, the former is still less steep than 
the latter in the region of low energy density.
As shown later, the EoSs for the QHT model Q4 and 
the quarkyonic model $\Lambda 380$ lead to the neutron-star
$MR$ curves consistent with the criterion Eq.(\ref{radii}).

In Fig.\ref{sound}, sound velocities are drawn as a function of $\rho_B$.
The dot-dashed curve is sound velocities in $\beta$-stable nucleonic matter.
Solid curves are those in quarkyonic matter for $\Lambda 380$, $\Lambda 350$ 
and $\Lambda 320$ with B1. 
There appear peak structures in the solid curves, being related to rapid 
increasing of pressures in the range of $2\rho_0\sim4\rho_0$. The dashed 
curve is sound velocities in the QHT matter with Q4 and the dotted one 
is those in $\beta$-stable hadronic matter with hyperon mixing, 
in which there appears no peak structure.
The dashed curve becomes $c_s >c$ in high-density region.
Also, the peak regions of solid curves become $c_s >c$, if B2 is used instead of B1
for nuclear parts. In such regions of $c_s >c$, 
sound velocities are approximated to be $c_s =c$.

It is interesting to notice that the peak structures in our quarkyonic-matter
results are somewhat similar to those for the QHC-matter EoS (QHC21) found
in \cite{KBH2022}. Our QHT-matter EoS gives no peak structure in sound velocities,
being different from both of them.

In the left panel of Fig.\ref{Pden2}, solid curves show pressures 
in quarkyonic matter for $\Lambda 380$ in the cases of using B1 and B2
for nuclear interactions, and short-dashed (dashed) curves are
partial pressures of neutrons (quarks) in respective cases.
The dot-dashed curve is pressures in $\beta$-stable nucleonic matter.
Pressures in quarkyonic matter are found to be completely dominated 
by neutron partial pressures. 
In order to reveal the reason why neutron pressures in quarkyonic matter
are far higher than those in $\beta$-stable nucleonic matter, we show the neutron 
chemical potentials in the cases of using B1 and B2 for nuclear interactions: 
In the right panel of Fig.\ref{Pden2}, neutron chemical potentials $\mu_n$ are 
drawn as a function of $\rho_B$. Lower and upper solid curves give neutron 
chemical potentials in quarkyonic matter for $\Lambda 380$ in the cases 
of using B1 and B2, respectively. The dot-dashed curve gives neutron chemical 
potential in $\beta$-stable nucleonic matter. 
The neutron chemical potentials in quarkyonic matter are far higher 
than those in the $\beta$-stable nucleonic matter, which makes neutron 
pressures in the former far higher than those in the latter. 
The reason of higher chemical potentials in the quarkyonic matter is 
because the existence of free quarks inside the Fermi sea gives nucleons 
extra kinetic energies by pushing them to higher momenta \cite{MR2019}.

\subsection{$MR$ diagrams}

We have the two types of hybrid EoSs,
the QHT-matter EoS and the quarkyonic-matter EoS. They are
combined with the $\beta$-stable nucleonic-matter EoS connected smoothly 
to the crust EoS \cite{Baym1,Baym2} in the low-density side.
The $MR$ relations of hybrid stars can be obtained by 
solving the TOV equations with these hybrid EoSs.

\begin{figure*}[ht]
\begin{center}
\includegraphics*[width=6.5in,height=3.5in]{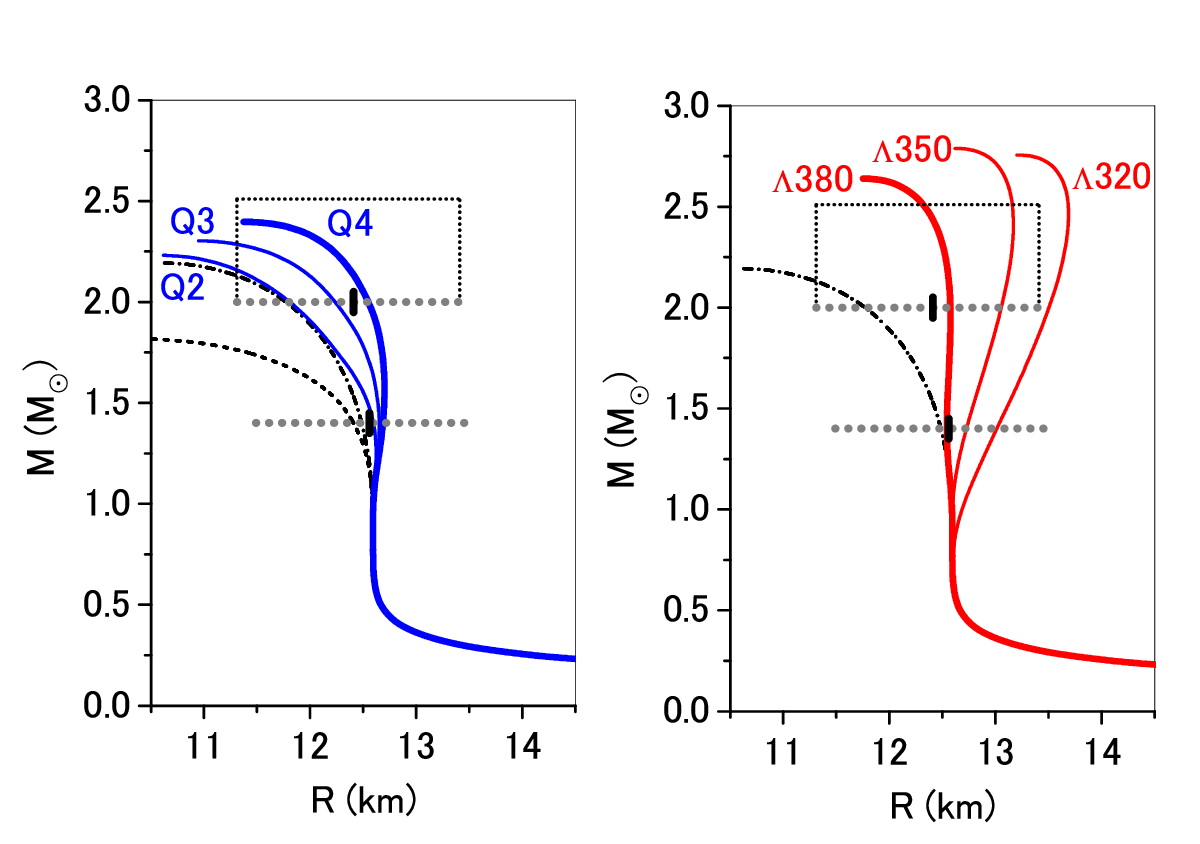}
\caption{(Color online) Star masses as a function of radius $R$.
The dot-dashed curves are by the $\beta$-stable nucleonic matter EoS. 
In the left panel, thin (thick) solid curves are by the QHT-matter EoSs
with Q2 and Q3 (Q4). The dotted curve is by the hadronic matter
EoS including hyperons.
In the right panel, thin (thick) solid curves are by the quarkyonic-matter
EoSs for $\Lambda 350$ and $\Lambda 320$ ($\Lambda 380$) with B1.
In both panels, the horizontal lines indicates 
$R_{1.4M_\odot}=12.56^{+1.00}_{-1.07}$ km and
$R_{2.0M_\odot}=12.41^{+1.00}_{-1.10}$ km, and the rectangle indicates 
the region of mass $M_{max}/M_\odot = 2.21^{+0.31}_{-0.21}$ \cite{Legred2021}.
}
\label{MR1}
\end{center}
\end{figure*}
%\end{figure}

In Fig.\ref{MR1}, star masses are given as a function of radius $R$.
The dot-dashed curves are obtained by the $\beta$-stable nucleonic matter EoS. 
In the left panel, thin (thick) solid curves are obtained by the QHT-matter EoSs 
with Q2 and Q3 (Q4). The dotted curve is by the hadronic-matter EoS 
including hyperons. In the cases of Q2, Q3 and Q4, the maximum masses 
are $M_{max}/M_\odot$= 2.23, 2.30, 2.40, respectively, and the radii 
at $2.0M_\odot$ are 11.8 km, 12.2 km, 12.5 km, respectively. 
In the right panel, thin (thick) solid curves are obtained by the 
quarkyonic-matter EoSs for $\Lambda 350$ and $\Lambda 320$ ($\Lambda 380$) 
with use B1 for nuclear interactions.
In the cases of $\Lambda 380$, $\Lambda 350$ and $\Lambda 320$,
the maximum masses are $M_{max}/M_\odot$= 2.64, 2.79, 2.76, respectively,
and the radii at $2.0M_\odot$ are 12.6 km, 13.1 km, 13.5 km, respectively. 
In both panels, the horizontal lines indicates 
$R_{1.4M_\odot}=12.56^{+1.00}_{-1.07}$ km and
$R_{2.0M_\odot}=12.41^{+1.00}_{-1.10}$ km, and the rectangle indicates 
the region of mass $M_{max}/M_\odot = 2.21^{+0.31}_{-0.21}$ \cite{Legred2021}.
%the {\bf {\large +}} mark points out
%the position ($2.1M_\odot$, 12.4 km).
The thick solid curve for Q4 in the left panel and that for $\Lambda 380$
in the right panel
%giving maximum mass $2.4M_\odot$ and $2.6M_\odot$, respectively. 
are found to be consistent with the criterion Eq.(\ref{radii}),
and the key features of $R_{2M_\odot} \approx R_{1.4M_\odot}$ are
found in these cases.

Then, it should be noted that the maximum mass $2.64M_\odot$ for
$\Lambda 380$ is substantially larger than the value $2.40M_\odot$ for Q4. 
The reason for such a difference between maximum masses can be understood
by comparing the $P(\rho_B)$ curves in Fig.\ref{Pden1}, where
the solid curve for $\Lambda 380$ increases more rapidly at onset of the 
quakyonic matter than the dashed curve for Q4 at onset of quark matter.
This means that the stiffness for former is larger than that for the latter.
In the case of QHT matter, it is not possible to obtain
such a rapid increasing of $P(\rho_B)$ in the low-density region, 
even if the $QQ$ repulsions are strengthened.

In the case of hadronic (nucleonic) matter, shown by the dotted (dot-dashed) 
curve in the left panel, the maximum mass is 1.82$M_\odot$ (2.19$M_\odot$).
The reduction of 0.37$M_\odot$ is due to the EoS softening
by hyperon (($\Lambda$ and $\Sigma^-$) mixing.
This softening is mainly caused by $\Sigma^-$ mixing:
If only $\Lambda$ mixing is taken into account, the maximum mass is
obtained as 2.06$M_\odot$ being close to the value of 2.19$M_\odot$
without hyperon mixing (dot-dashed curve). 
Thus, massive stars with $M>2M_{\odot}$ cannot be obtained by
the hadronic matter EoSs with hyperon ($\Lambda$ and $\Sigma^-$) 
mixing  \cite{YFYR14,YFYR16,YTTFYR17}. 
On the other hand, the value of $R_{1.4M_\odot}$ is 12.4 (12.5) km
in the case of hadronic (nucleonic) matter, which means that
the hyperon mixing does not depend much on $R_{1.4M_\odot}$.

%On the other hand, the $MR$ curves for the QHT-matter EoS with Q4 
%and the quakyonic matter EoS with $\Lambda 380$, being consistent with 
%Eq.(\ref{radii}), give quite large values of maximum masses $2.4M_\odot$ 
%and $2.6M_\odot$, respectively.

\begin{figure}[ht]
\begin{center}
\includegraphics*[width=8.6cm]{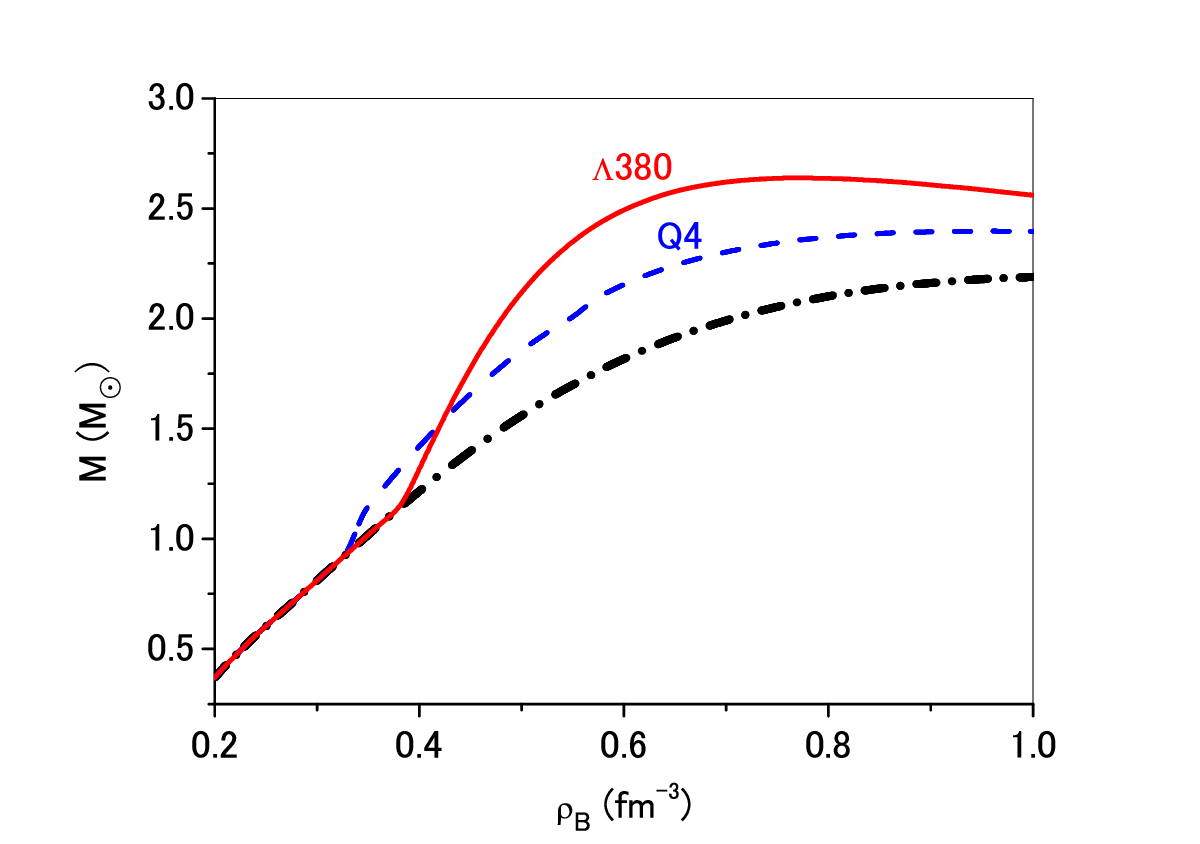}
\caption{(Color online) Star masses as a function of central baryon density $\rho_{Bc}$.
The dot-dashed curves are by the $\beta$-stable nucleonic EoS. 
The solid curve is by the quarkyonic-matter EoS for $\Lambda 380$ in the case of using B1.
The short-dashed curve is by the QHT-matter EoS for Q4.}
\label{Mdenc}
\end{center}
\end{figure}

In Fig.\ref{Mdenc}, star masses are given as a function 
of central baryon density $\rho_{Bc}$.
The dot-dashed curves are by the $\beta$-stable nucleonic matter EoS. 
The solid curve is obtained by the quarkyonic-matter EoS for $\Lambda 380$ with B1, 
and the dashed curve is by the QHT-matter EoS for Q4, where the onset density
in the former (latter) 0.39 (0.33) fm$^{-3}$.
Both of them are consistent with Eq.(\ref{radii}), but the former mass curve 
for $\rho_{Bc}$ is considerably above the latter one, 
as well as the corresponding $MR$ curves.

\begin{figure}[ht]
\begin{center}
\includegraphics*[width=8.6cm]{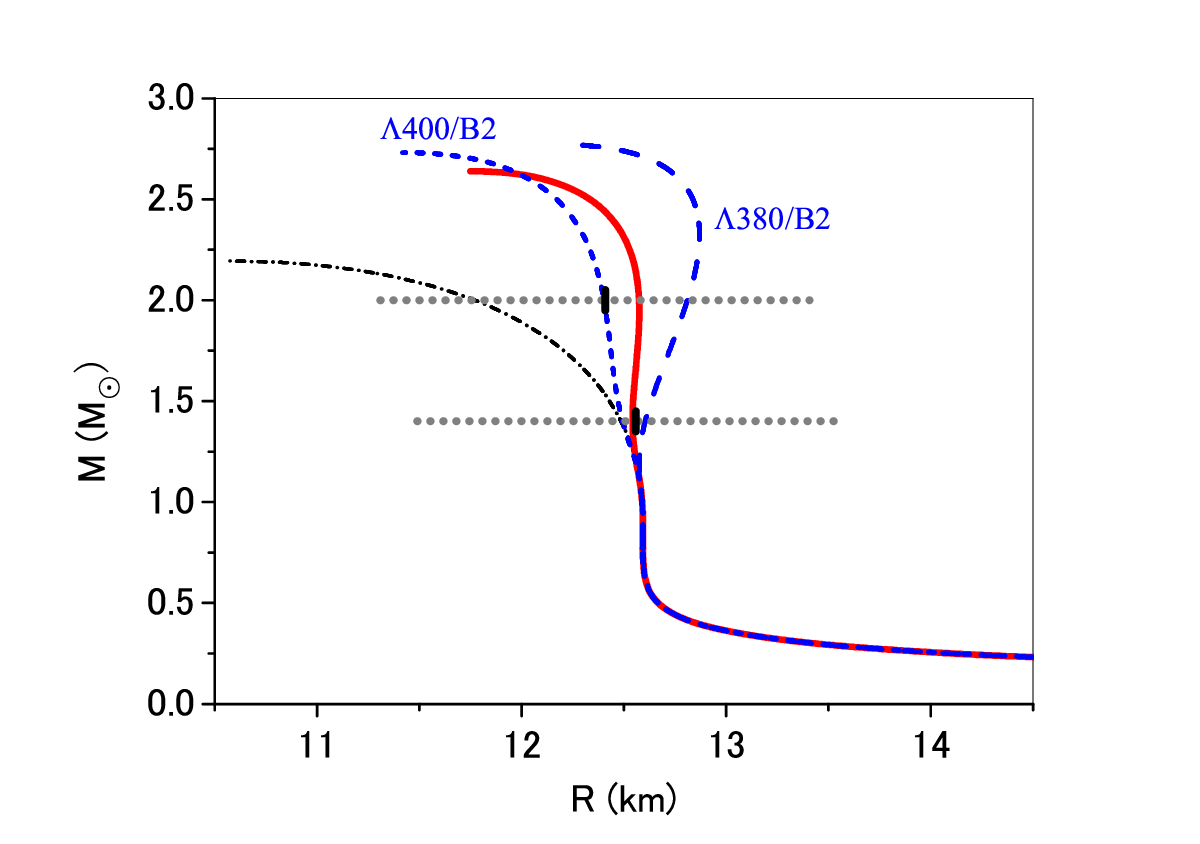}
\caption{(Color online) Star masses as a function of radius $R$.
The dot-dashed curves are by the $\beta$-stable nucleonic matter EoS.
The solid curve is for $\Lambda 380$ with B1.
Dashed and short-dashed curves are by the quarkyonic-matter EoSs for 
$\Lambda 380$ and $\Lambda 400$ with B2, respectively.
The horizontal dotted lines indicates $R_{1.4M_\odot}=12.56^{+1.00}_{-1.07}$ km 
and $R_{2.0M_\odot}=12.41^{+1.00}_{-1.10}$ km.}
%The {\bf {\large +}} mark points out the position ($2.1M_\odot$, 12.4 km).}
\label{MR2}
\end{center}
\end{figure}

In Fig.\ref{MR2}, star masses are given as a function of radius $R$.
The solid curve is obtained by the quarkyonic-matter EoS for 
$\Lambda 380$ with use of B1 ($V_{nn}$) for nuclear interactions, 
given also in Fig.\ref{MR1}. Dashed and short-dashed curves are by the 
quarkyonic-matter EoSs for $\Lambda 380$ and $\Lambda 400$, respectively, 
in the case of using B2 ($V_{nn}$+$V_{nnn}$) instead of B1.
The difference between solid and dashed curves demonstrates
the effect of the three-neutron repulsion $V_{nnn}$, giving the larger 
maximum mass and larger value of $R_{2.0M_\odot}$. The short-dashed curve for
$\Lambda 400$ indicates that this effect of $V_{nnn}$ to increase mass 
and radius is cancelled out by taking larger values of $\Lambda$.

\begin{figure}[ht]
\begin{center}
\includegraphics*[width=8.6cm]{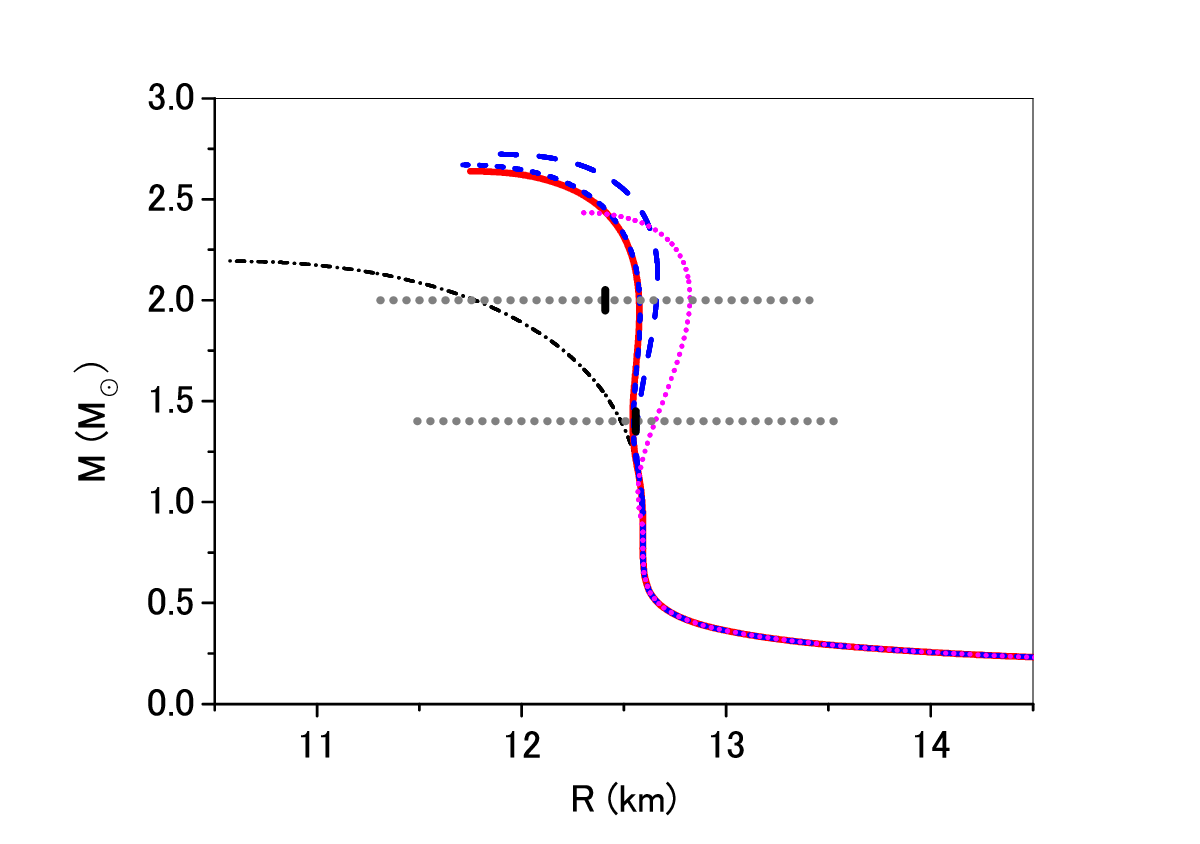}
\caption{(Color online) Star masses as a function of radius $R$.
The dot-dashed curves are by the $\beta$-stable nucleonic matter EoS.
The solid curve is obtained by the quarkyonic-matter EoS
for $\Lambda 380$ with $\kappa=0.3$ in the case of using B1.
The dashed (short-dashed) curve is for $\Lambda 380$ with $\kappa=0.3$ 
($\kappa=0.4$) by the approximation to neglect potential sectors
in quark energy densities.
The dotted curve is obtained by replacing the potential energy density 
in Eq.(\ref{edenn}) to Eq.(\ref{vden}).
The horizontal lines indicates $R_{1.4M_\odot}=12.56^{+1.00}_{-1.07}$ km 
and $R_{2.0M_\odot}=12.41^{+1.00}_{-1.10}$ km.}
\label{MR3}
\end{center}
\end{figure}

In Fig.\ref{MR3}, star masses are given as a function of radius $R$.
The solid curve is obtained by the quarkyonic-matter EoS for 
$\Lambda 380$ with $\kappa=0.3$ in the case of using B1, 
given also in Fig.\ref{MR1}.
The dashed curve is obtained by the approximation used in \cite{MR2019},
where the $QQ$ interactions are neglected and the quark energy density 
Eq.(\ref{edenq}) is replaced by the kinetic energy density.
Then, the difference between short-dashed and dashed curves is
due to this approximation. The short-dashed curve is obtained 
by taking $\kappa=0.4$ under this approximation.
The similarity between solid and short-dashed curves means that 
the deviation due to this approximation is canceled out 
by adjusting the value of $\kappa$.
In the same case of $\Lambda 380$ and $\kappa=0.3$ with B1,
the dotted curve is obtained by replacing the potential energy density 
in Eq.(\ref{edenn}) to Eq.(\ref{vden}), being the approximated treatment
in \cite{MR2019}. This approximation to use Eq.(\ref{vden}) is found 
to reduce masses and to increase radii. 

\subsection{Discussion}

In \cite{Legred2021}, they present the neutron-star properties
such as maximum mass, radius, tidal deformability,
pressure and central density inferred from their analysis,
for which the median and $90\%$ highest-probability-density credible regions 
are given.  From Table II of \cite{Legred2021}, we choose the quantities 
in the case of w/J0740+6620 Miller+ in order to compare with the corresponding 
values obtained from our QHT-matter and the quarkyonic matter EoSs.
In Table \ref{observable}, tabulated are maximum masses $M_{max}$, 
pressures $p$ at $\rho_0$, $2\rho_0$ and $6\rho_0$, radii $R$ and
dimensionless tidal deformabilities $\Lambda$ at $1.4M_\odot$ and $2.0M_\odot$,
central densities $\rho_c$ at $1.4M_\odot$, $2.0M_\odot$ and $M_{max}$.
Here, our results are for the $\beta$-stable nucleonic matter EoS
denoted as NUC, the QHT-matter EoS Q4 and the quarkyonic matter EoS V380.
These EoSs are adjusted so as to reproduce $R_{1.4M_\odot}$
with an accuracy of a few hundred meters.
Then, the key feature of $R_{2M_\odot} \approx R_{1.4M_\odot}$ is 
found in the cases of Q4 and V380 EoSs, contrastively to the case
of the nucleonic EoS giving $R_{2M_\odot} < R_{1.4M_\odot}$.
The values of $R_{2.0M_\odot}$, central densities and tidal deformabilities 
for Q4 and V380 EoSs are far closer to the median values than those for 
nucleonic EoS, demonstrating the clear impacts of quark phases
in Q4 and V380 EoSs.
The deviations from the median values in the latter are
considerably larger than those in the formers. 
Especially, the values of $\Lambda_{1.4}$ and $\Lambda_{2.0}$ 
for the nucleonic EoS are noted to be out of $90\%$ credible regions.

In the case of the quarkyonic matter EoS for V380,
the values of $M_{max}$ and $p(6\rho_0)$ are found to be far larger than
that for the nucleonic EoS. It is interesting that such a large value of 
$M_{max}$ can be obtained straightforwardly from the quarkyonic-matter EoS,
considering the implication of the large mass $(2.35 \pm 0.17)M_\odot$
for PSR J0952-0607 \cite{Romani2022}.
The reason why a large value of $M_{max}$ is obtained n the case of 
the quarkyonic matter EoS is because the pressure rises rapidly 
in the region of $\rho_B \sim 2\rho_0$ as found in Fig.\ref{Pden1}.
In the McLerran-Reddy model of the quarkyonic matter,
the resulting EoS is mainly controlled by the one parameter $\Delta_{qyc}$
for Fermi-layer thickness. Then, it is difficult to reproduce 
simultaneously $M_{max}=2.2M_\odot$ and $R_{2.0M_\odot}=12.4$ km.

\begin{table}
\begin{center}
\caption{Maximum masses $M_{max}$, pressures $p$ at $\rho_0$, $2\rho_0$ 
and $6\rho_0)$, radii $R$ and tidal deformabilities $\Lambda$ at 
$1.4M_\odot$ and $2.0M_\odot$, central densities $\rho_c$ at $1.4M_\odot$, 
$2.0M_\odot$ and $M_{max}$.
Results for the $\beta$-stable nucleonic matter EoS denoted as NUC, 
the QHT-matter EoS Q4 and the quarkyonic matter EoS V380 are
compared with the values taken from \cite{Legred2021}.
}
\label{observable}
\vskip 0.2cm
\begin{tabular}{|c|cccc|}\hline
      & \ NUC \ & \ \  Q4 \ \  & \  $\Lambda 380$ \ & Ref.\cite{Legred2021} \\
\hline
$M_{max}/M_\odot$   & 2.19 &  2.40 &  2.64  &  $2.21^{+0.31}_{-0.21}$  \\
$p(\rho_0)$ ($10^{33} {\rm dyn/cm^2}$) &  5.27 & 5.27 & 5.27 & $4.30^{+3.37}_{-3.80}$ \\
$p(2\rho_0)$ ($10^{34} {\rm dyn/cm^2}$) & 2.76 & 5.09 & 4.42 & $4.38^{+2.46}_{-2.96}$ \\
$p(6\rho_0)$ ($10^{35} {\rm dyn/cm^2}$) & 6.94 & 12.0 & 22.6 & $7.41^{+5.87}_{-4.18}$ \\
$R_{1.4M_\odot}$ (km) &  12.5 &  12.7 &   12.5    &  $12.56^{+1.00}_{-1.07}$  \\
$R_{2.0M_\odot}$ (km) &  11.8 &  12.5 &   12.6    &  $12.41^{+1.00}_{-1.10}$  \\
$R_{2.0M_\odot}-R_{1.4M_\odot}$ (km) & $-0.72$ & $-0.14$ & $+0.03$ & $-0.12^{+0.83}_{-0.85}$ \\
$\Lambda_{1.4}$  &  779  &   525   &   473     & $507^{+234}_{-242}$   \\
$\Lambda_{2.0}$  &  128  &   46    &   49      & $44^{+34}_{-30}$   \\
$\rho_c(1.4M_\odot)$ ($10^{14} {\rm g/cm^3}$) & 7.9 &  6.6  &   6.8   &  $6.7^{+1.7}_{-1.3}$  \\
$\rho_c(2.0M_\odot)$ ($10^{14} {\rm g/cm^3}$) & 12. &  9.1  &   8.0   &  $9.7^{+3.6}_{-3.1}$  \\
$\rho_c(M_{max})$ ($10^{15} {\rm g/cm^3}$) &  1.8   &  1.6  &   1.3   &  $1.5^{+0.3}_{-0.4}$  \\
\hline
\end{tabular}
\end{center}
\end{table}

\section{Conclusion}

The observed masses and radii of neutron stars give constraints
on the dense matter EoSs and resulting $MR$ diagrams.
In this sense, the observations of massive stars over $2M_{\odot}$ and 
the NICER implication of $R_{2M_\odot} \approx R_{1.4M_\odot}$
are critically important for restricting neutron-star matter EoSs.
In the case of hadronic matter, even if the nucleonic matter EoS is constructed 
so as to be stiff enough to give the maximum mass over $2M_{\odot}$, the 
hyperon mixing brings about a remarkable softening of the EoS.
The EoS-softening by hyperon mixing can be reduced, for instance,
by introducing many-body repulsions which work universally for 
every kind of baryons. However, such a repulsive effect does not 
cancel out completely the EoS softening by hyperon mixing:
In the case of hadronic matter EoS with hyperon mixing, it is
difficult to obtain maximum masses over $2M_{\odot}$.
The most promising approach to solve this ``hyperon puzzle" is to 
assume the existence of quark phases in inner cores of neutron stars,
namely hybrid stars having quark matter in their cores. 

When quark deconfinement phase transitions from a hadronic-matter EoS 
to a sufficiently stiff quark-matter EoS are taken into account in 
the neutron-star interiors, repulsive effects such as $QQ$ repulsions 
in quark phases are needed in order to obtain sufficiently stiff EoSs
resulting in massive hybrid stars with masses over $2M_{\odot}$.
In our QHT matter, it is possible to reproduce maximum masses
over $2M_{\odot}$ consistently with the NICER implication, where
the $QQ$ repulsion is taken to be strong enough and the quark-hadron 
transition density is adjusted so as to be about $2\rho_0$ by tuning 
of the density dependence of effective quark mass.

In the quarkyonic matter, the degrees of freedom inside the Fermi sea 
are treated as quarks, and nucleons exist at the surface of the Fermi sea.
The existence of free quarks inside the Fermi sea gives nucleons 
extra kinetic energy by pushing them to higher momenta.
This mechanism of increasing pressure is completely different from 
the above mechanism of EoS stiffening by strong $QQ$ repulsions in the QHT matter.
In calculations of $MR$ diagrams with the quarkyonic-matter EoS,
the critical quantity is the thickness $\Delta_{qyc}$ of Fermi layer
controlled by the parameters $\Lambda$ and $\kappa$.
With the reasonable choice of these parameters, the $MR$ curves of
quarkyonic hybrid stars are obtained so as to be consistent 
with the NICER implication.

As well as $R_{2.0M_\odot}$, central densities and tidal deformabilities are 
inferred from the analysis of the NICER data. The QHT-matter and quarkyonic EoSs
can be adjusted so as to reproduce these inferred quantities far closer 
to the median values than those for nucleonic matter EoS, 
demonstrating the clear impacts of quark phases in these cases..

Thus, the reasonable  $MR$ curves of neutron stars can be derived from 
both QHT-matter and quarkyonic-matter EoSs, having completely different 
mechanisms to stiffen EoSs.
However, when both EoSs are adjusted so as to be consistent with the NICER
implication, the maximum mass for the quakyonic-matter EoS is considerably 
larger than that for the QHT-matter EoS.

\section*{Acknowledgments}
%\acknowledgment
{The authors would like to thank D. Blaschke 
for valuable comments and fruitful discussions.}

\end{document}